\documentclass[a4paper,hyphens]{article}
\usepackage[margin=1in]{geometry} 

\usepackage{amsmath}
\usepackage{amsthm}
\usepackage{amssymb}

\usepackage[utf8]{inputenc}
\usepackage{hyperref}
\hypersetup{
	unicode,
	pdftitle={Targeted synthesis of vanadium dioxide thin films}
        breaklinks=true
}

\usepackage[sort&compress,numbers,square]{natbib}
\usepackage{graphicx, color, soul}
\usepackage[dvipsnames]{xcolor,colortbl}
\definecolor{Gray}{gray}{0.85}
\usepackage{upgreek}

\title{Targeted synthesis of polycrystalline vanadium dioxide thin films via post-deposition annealing}
\author{Kirill Trunov$^1$, Yuri Lebedinskii$^2$, Ilya Zavidovskiy$^3$, Sergey Novikov$^3$, Alexander\\Morozov$^{3,4}$, Petr Shvets$^5$, Ksenia Maksimova$^5$, Andrei Zenkevich$^1$, Anton Khanas$^{1*}$}

\date{\small{\textit{
$^1$Lab of Functional Materials and Devices for Nanoelectronics, Moscow Institute of Physics and Technology (National Research University), Institutsky lane 9, Dolgoprudny, Moscow region, 141700, Russian Federation\\
$^2$Center of Shared Reasearch Facilities, Moscow Institute of Physics and Technology (National Research University), Institutsky lane 9, Dolgoprudny, Moscow region, 141700, Russian Federation\\
$^3$Center for Photonics and 2D Materials, Moscow Institute of Physics and Technology (National Research University), Institutsky lane 9, Dolgoprudny, Moscow region, 141700, Russian Federation\\
$^4$Programmable Functional Materials Lab, Center for Neurophysics and Neuromorphic Technologies, Moscow, 127495, Russian Federation\\
$^5$Research and Educational Center “Functional Nanomaterials”, Immanuel Kant Baltic Federal University, Aleksandra Nevskogo 14, Kaliningrad, 236041, Russian Federation\\}}
$^*$Author, to whom correspondence should be addressed: \texttt{khanas@phystech.edu}
}

\begin{document}

\maketitle

\begin{abstract}

{\noindent}Implementation of neuromorphic hardware is a promising way to improve the computing efficiency and decrease the energy consumption of artificial neural networks. For this purpose, electronic elements emulating the behavior of synapses and neurons have to be developed. In order to realize electronic artificial neurons, threshold resistive switches or memristors can be efficiently used. One of the most widespread materials for threshold switches is vanadium dioxide due to its property to demonstrate the metal-insulator transition at a temperature about $70~^\circ$C. However, the processes of VO$_2$ synthesis are quite restrictive in temperature and gas atmosphere conditions, which hinders its integration into CMOS fabrication. In this work, we propose a new method of VO$_2$ synthesis: reactive pulsed laser deposition from metallic V target in oxygen atmosphere at room temperature, followed by vacuum annealing. Our method enables target synthesis of an appropriate VO$_2$ phase in a polycrystalline thin film form by finely tuning oxygen pressure during room temperature deposition, which allows to relax the equipment demands, such as high temperature heating in oxygen. Successful targeted VO$_2$ synthesis under fabrication conditions close to back-end-of-line CMOS production, achieved in this work, show the way toward its large-scale microelectronic integration for neuromorphic hardware creation.
\end{abstract}


\newpage
\section{Introduction}\label{s:introduction}

\hspace{12pt} Artificial neural network (ANN) algorithms have attracted significant attention in recent years due to their outstanding performance in image and voice recognition, data analysis and content generation tasks. The largest drawback in this field, hindering its progress, is enormous power consumption \cite{kindig_ai_nodate}, arising from poor coherence of software and hardware on the level of system architecture. In order to overcome this issue, development of novel electronic components, which could emulate the behavior of neurons and synapses on various levels of realism (depending on ANN type), is crucial \cite{zhu_comprehensive_2020}.

Threshold resistive switches, also known as volatile threshold memristors, are among the most applicable elements for implementation of compact artificial neurons (neuristors) \cite{han2022}. They can be realized using various physical mechanisms: metal-insulator transition (MIT), formation of volatile metallic filaments in dielectric matrices, single-transistor latching \cite{han2022}, or ignition of glow discharge \cite{goool}. MIT-based resistive switches have significant advantages, such as high switching speed, low transition energy, high reliability in term number of switching cycles, good scalability and a possibility of CMOS integration \cite{pickett_sub-100_2012, li_crossmodal_2024}.

Among the variety of materials that exhibit MIT, vanadium oxides (VO$_x$) are particularly interesting. VO$_x$ may exist in a great number of phases with different stoichiometric composition and physical properties, many of which undergoes MIT \cite{hu_vanadium_2023}. However, only for VO$_2$ and V$_3$O$_5$ the transition occurs above room temperature (RT), so only these phases are suitable for practical implementation of threshold switches. Vanadium dioxide is relatively more frequently studied in the context of neuromorphic electronics due to lower transition temperature (about $67~^\circ$C, in comparison to about $180~^\circ$C for V$_3$O$_5$), which enables lower energy consumption per switching.
 
Vanadium dioxide thin films can be grown using various methods: magnetron sputtering \cite{tsai2003properties, yang2018controlling, yi2018biological, lee_controlling_2021, bidoul2024tuning}, atomic layer deposition (ALD) \cite{varini_pulsed_2025, peng2025ald, kapoguzov2025formation}, chemical vapor deposition (CVD) \cite{RAJESWARAN2020122230}, molecular beam epitaxy (MBE) \cite{rata2004growth}, cathode arc sputtering \cite{shvets_review_2019, shvets_raman_2024}, sol-gel \cite{ningyi_valence_2002} or pulsed laser deposition (PLD). PLD has several advantages: high energy of depositing species (of the order of $100~$eV) \cite{eason_pulsed_2006} enables surface diffusion of adatoms, thus facilitating epitaxial VO$_2$ growth on Al$_2$O$_3$ or TiO$_2$ substrates \cite{kim1994pulsed, hattori_investigation_2020, zhou_revealing_2021, kaydashev_reconfigurable_2023, atul_strong_2024}, which yields highest quality films in comparison to other synthesis methods. Moreover, PLD can yield high-quality VO$_2$ films on SiO$_2$/Si substrates \cite{kutepov_optimizing_2024, varini_pulsed_2025}, which are more preferable for CMOS-compatible processes. However, a major problem in the implementation of VO$_2$ into microelectronic technology is the simultaneous necessity of high temperature ($400$-$500~^\circ$C) and oxygen ambient during deposition with precise parameter control. One of the ways to evade these complications is post-deposition annealing in either vacuum or oxygen ambient, which can be employed for crystallization of amorphous VO$_x$ films \cite{peng2025ald}, for modification of electrical properties of VO$_2$ \cite{varini_pulsed_2025} or for solid-phase transformation of V$_2$O$_5$ \cite{ningyi_valence_2002} or V$_2$O$_3$ \cite{lee_controlling_2021} phases into VO$_2$. However, these processes still require a preliminary stoichiometric matching or crystallization of the sample.

In this paper, we introduce a two-step method of VO$_2$ growth via reactive PLD from metallic V target at RT in the oxygen atmosphere, followed by the post-deposition vacuum annealing. This method enables good control of the VO$_x$ phase composition, enough to ensure the single phase VO$_2$ synthesis in a polycrystalline thin film form. Reduction of the as grown amorphous VO$_x$ films (transformation of V$^{5+}$ to V$^{4+}$ ion oxidation state, associated with the VO$_2$ phase), is confirmed via X-ray photoelectron spectroscopy with \textit{in situ} vacuum heating. To verify the crystalline phase properties of the films, we employ X-ray diffraction, Raman spectroscopy and electrical measurements, which allow to optimize the oxygen pressure during growth for precise phase composition tuning. Electrical measurements in lateral geometry, demonstrating low-voltage threshold switching, prove the eligibility of obtained VO$_2$ films for neuromorphic applications. Additionally, the proposed method allows to avoid high temperature heating in oxygen atmosphere at any synthesis step, which relaxes the requirements for fabrication equipment, thus showing the possible way for truly CMOS-compatible integration of VO$_2$ thin films.

\section{Experimental methods}\label{s:experimental}

\hspace{12pt} \textbf{Synthesis method.} Vanadium oxide thin films were grown on a Si(100) substrate with a 240-nm-thick thermal SiO$_2$ layer via pulsed laser deposition (PLD). Deposition was performed with a Q-switched Nd:YAG laser with a laser beam wavelength of $1064~$nm, $10~$Hz repetition rate and $\sim$10~ns pulse duration. Diameter of a focused beam did not exceed $1~$mm. Growth was carried out using the metallic V target in oxygen with pressure varied in the range of $1$$-$$3~$Pa. The substrate was kept at RT during growth. The films were grown using 10,000 pulses with energy of $\sim 245~$mJ. After deposition, the samples were annealed in the same vacuum chamber in high vacuum (pressure $< 4 \times 10^{-6}~$mbar). Samples' temperature was increased for $\sim 5~$min up to the annealing temperature of $490 \pm 10~^\circ$C (verified by infrared optical pyrometer). Total annealing time at peak temperature equaled $20~$min.

\textbf{X-ray photoelectron spectroscopy.} X-ray photoelectron spectroscopy (XPS) was performed in the high vacuum using a Theta Probe spectrometer with an Al K$_\alpha$ X-ray source ($1486.6~$eV), equipped with a monochromator. To perform measurements with \textit{in situ} vacuum heating, a self-developed heater was utilized with thermocouple temperature control inside the spectrometer vacuum chamber.

\textbf{X-ray diffraction.} X-ray diffraction (XRD) patterns were obtained in grazing incidence configuration (fixed X-ray angle of incidence $\omega = 2^{\circ}$, with respect to the sample surface, with varying detector angle) with Thermo ARL X'TRA diffractometer with Cu $K_\alpha$ radiation. 

\textbf{Raman spectroscopy.} Raman spectra were obtained with a LabRAM HR Evolution Raman spectrometer using an excitation wavelength of $532~$nm. 

\textbf{Electrodes fabrication.} Lateral electrodes for electrical measurements (both in Van Der Pauw and two probe configurations) were fabricated using PLD-grown Ru, patterned via UV optical lithography and lift-off process. For $I$$-$$V$ measurements, a gap between electrodes equaled $3~\upmu$m.

\textbf{Electrical measurements.} DC $I(V)$ characteristics of Ru/VO$_2$/Ru lateral devices were measured using Cascade Microtech Summit 11000M probe station equipped with Keysight B1500A semiconductor parameter analyzer. Temperature dependence of film resistivity was measured in the Van Der Pauw geometry with a Quantum Design Physical Property Measurement System (PPMS).

\textbf{Stylus profilometry.} Ambios XP-Plus 200 Stylus Profiler was used to determine the thickness of the films, which was about $70~$nm in all experiments.

\section{Results and discussion}\label{s:results}

\subsection[Annealing-induced reduction and crystallization of amorphous VOx films]{Annealing-induced reduction and crystallization of amorphous VO$_x$ films}

\begin{figure}[b!]
\begin{center}
\includegraphics[width=0.45\linewidth]{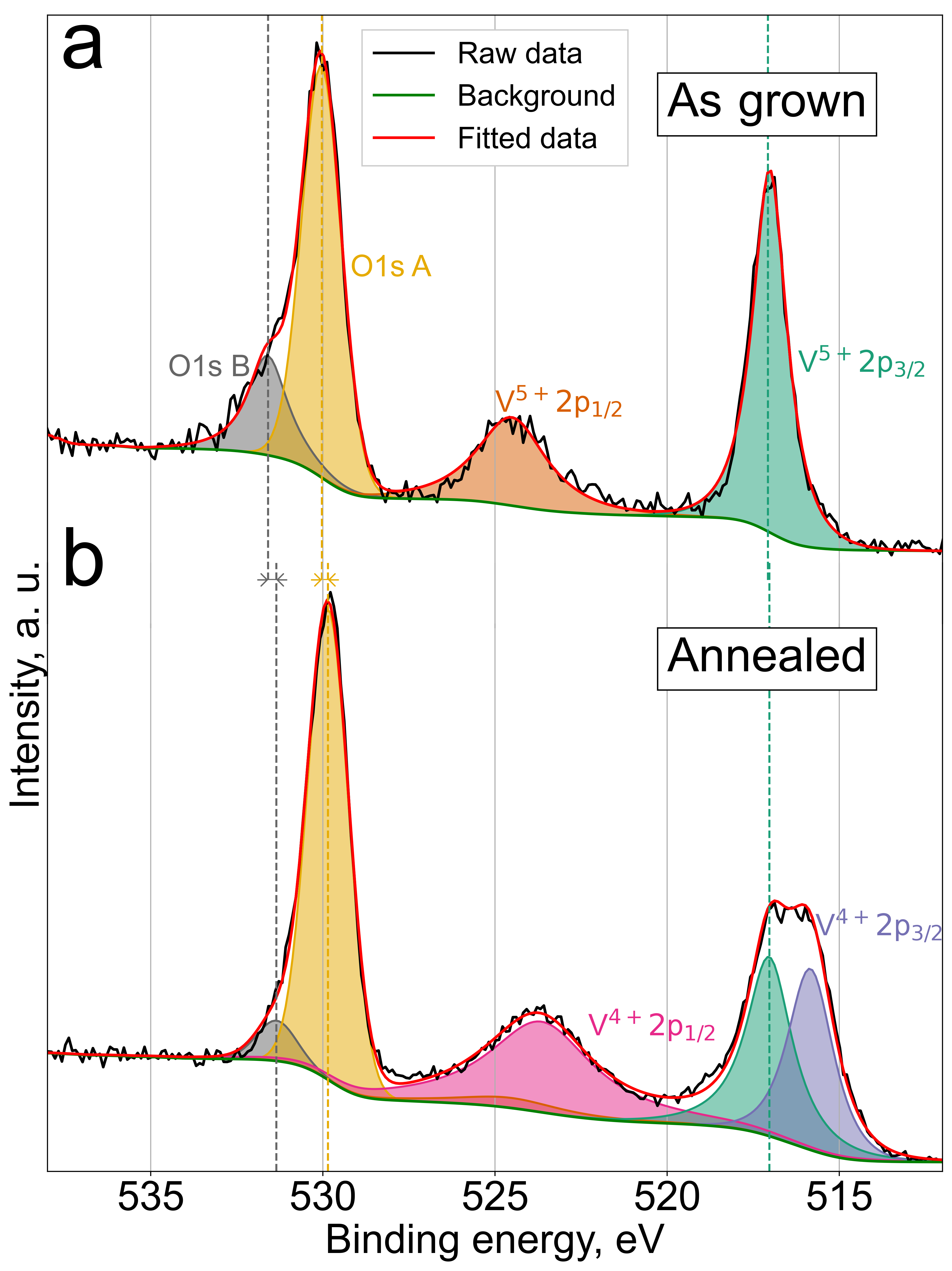}
\caption{X-ray photoelectron spectra of a VO$_x$ film after RT deposition (a) and after \textit{in situ} vacuum annealing at 500$~^\circ$C (b). Both measurements were performed at RT.}
\label{xps}
\end{center}
\end{figure}

To study the effect of annealing on amorphous VO$_x$ thin films grown at RT, an XPS experiment was conducted with \textit{in situ} vacuum heating (\textbf{fig.\ref{xps}}). As grown film was exposed to the atmosphere and then put into an XPS vacuum chamber. Fig.\ref{xps}a represents a spectrum of the film in the as received condition. The spectrum fitting reveals the presence of two oxygen (O$1s$) peaks: the one with lower binding energy, about 530 eV (denoted on the figure as "A"), is attributed to lattice oxygen, while the higher binding energy one ("B") is attributed to the surface contamination (such as water). The surface layers of the film (several nm thick -- XPS probing depth) are in the form of the highest vanadium oxide, V$_2$O$_5$, determined from the V$2p_{3/2}$ peak position relative to the lattice oxygen O$1s$ peak (separation of 13 eV), which allows to ascribe the V doublet to V$^{5+}$ oxidation state \cite{silversmit2004determination}. Annealing the sample at 500$~^\circ$C in a high vacuum (pressure $< 10^{-7}~$mbar) has multiple effects (fig.\ref{xps}b): i) the oxygen peak, associated with surface contamination is significantly reduced, however, not completely; ii) a second V doublet appears at lower binding energies, while the initial V$^{5+}$ peaks are reduced in intensity. The energy separation between O$1s$ and an emerged V peak (V$2p_{3/2}$ component) equals $13.6~$eV, closely matching the reference for V$^{4+}$ 2$p_{3/2}$ case \cite{silversmit2004determination, koethe2006bulk}. These changes can be explained by the reduction of the VO$_x$ film in the surface layer through the exit of oxygen to the vacuum \cite{silversmit2004determination}.

\subsection{Oxygen pressure optimization}

\begin{figure}[t!]
\centering
\includegraphics[width=0.85\linewidth]{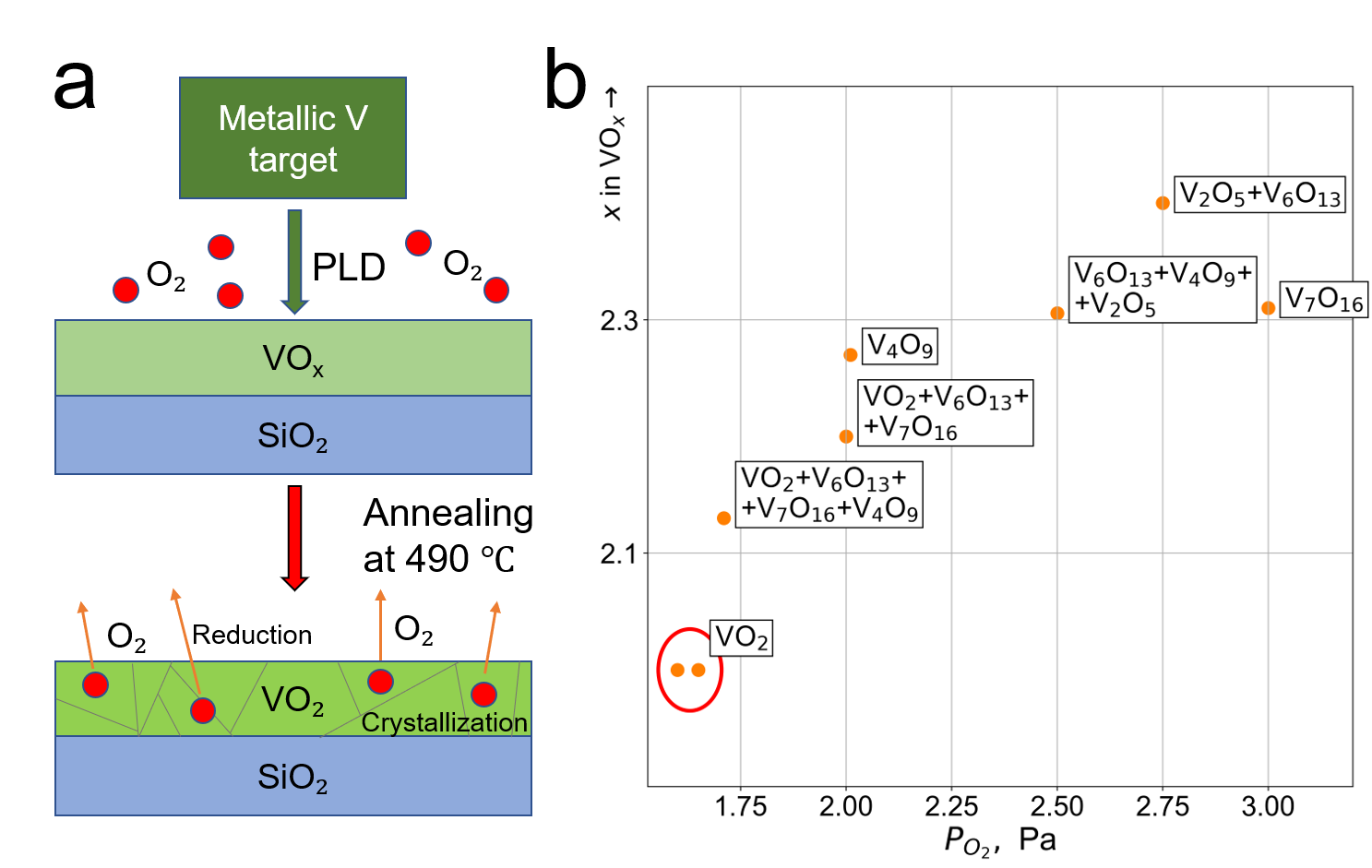}
\caption{a) Schematic of reactive PLD of amorphous VO$_x$ film and subsequent vacuum annealing at 490$~^\circ$C. b) Phase composition of VO$_x$ samples as a function of oxygen pressure during deposition. Phase content in the films was established qualitatively using XRD and Raman spectroscopy. Optimal parameters region for VO$_2$ synthesis is highlighted with a red ellipse.}
\label{phases_plot}
\end{figure}

\hspace{12pt} The XPS \textit{in situ} reduction experiment shows the principal possibility to obtain a precise vanadium oxide phase -- in our case, VO$_2$ -- by separate RT reactive deposition in oxygen and post-deposition vacuum annealing (not necessarily performed in the same chamber without air exposure, since the XPS experiment was performed \textit{after} air exposure). In order to find the optimal deposition conditions for targeted growth of VO$_2$ thin films, a series of samples was prepared following the proposed procedure (\textbf{fig.\ref{phases_plot}a}). Oxygen pressure during deposition was varied in the range of $1$ -- $3~$Pa. The conditions of post-deposition annealing were kept the same for all samples: $490\pm10~^{\circ}$C, chosen due to literature references on characteristic temperature for VO$_2$ phase growth \cite{ningyi_valence_2002, lee_controlling_2021}, for 20 min total at peak temperature with $\sim$5 min heating from RT to peak.

\begin{figure}[t!]
\begin{center}
\includegraphics[width=\linewidth]{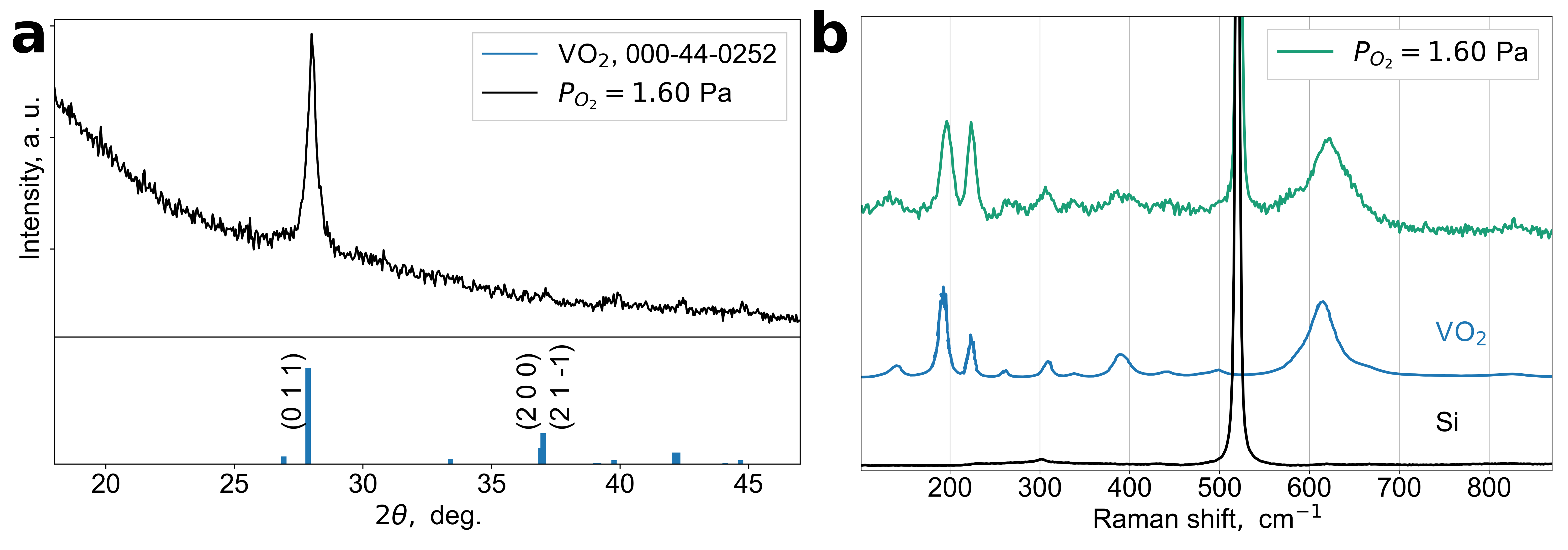}
\caption{a) XRD pattern of a sample grown with $1.60~$Pa O$_2$ pressure and attributed to a VO$_2$ phase. b) Raman spectrum of the same film.}
\label{vo2}
\end{center}
\end{figure}

It was found that the reference as grown VO$_x$ samples without vacuum annealing do not show any diffraction or Raman peaks and, therefore, are considered to be amorphous. The annealed samples demonstrate the diffraction peaks, associated with crystalline order, and the resulting crystalline phase content strongly depends on the oxygen pressure during deposition: as expected, lower oxygen pressure leads to the less oxidized crystalline phase of VO$_x$. Raman spectroscopy confirms and complements the XRD analysis -- the spectra can be analyzed by the presence of "fingerprints" of pure crystalline phases \cite{shvets_review_2019}, while, similarly to XRD, amorphous samples demonstrate only wide halos on the Raman spectra, the analysis of which in the case of polyphase samples is challenging \cite{shvets_correlation_2021}. Due to grazing incidence XRD geometry, partial overlap of the diffraction and Raman peaks of different phases and polycrystalline character of the thin films (which distorts the reference peak amplitude ratios), the quantitative determination of the phase content was severely complicated, so we concentrated on the search for "pure" VO$_2$ (at least, not showing the signs of presence of other crystalline phases) and identified the presence of other phases in the samples, grown at different oxygen pressures, only qualitatively (fig.\ref{phases_plot}). Namely, on the way of reducing the oxygen pressure during deposition, we have found the characteristic XRD and Raman peaks, corresponding to V$_2$O$_5$, V$_7$O$_{16}$, V$_4$O$_9$ and V$_6$O$_{13}$ (\textbf{fig.S1, S2}, Supporting Information). The optimal oxygen pressure during PLD, which allows to obtain only VO$_2$ both in XRD patterns and Raman spectra, equals $1.60$$-$$1.65~$Pa (\textbf{fig.\ref{vo2}}). Note that VO$_2$ was the least oxidized crystalline phase we were able to obtain, since the samples grown at less than $1.5~$Pa O$_2$ pressure did not demonstrate any peaks on XRD patterns and Raman spectra, and were considered to be amorphous. The latter observation also does not allow to rule out the presence of amorphous inclusions in the supposedly crystalline samples as well.

\subsection[VO2 film electrical properties]{VO$_2$ film electrical properties}

\begin{figure}[b!]
\begin{center}
\includegraphics[width=0.9\linewidth]{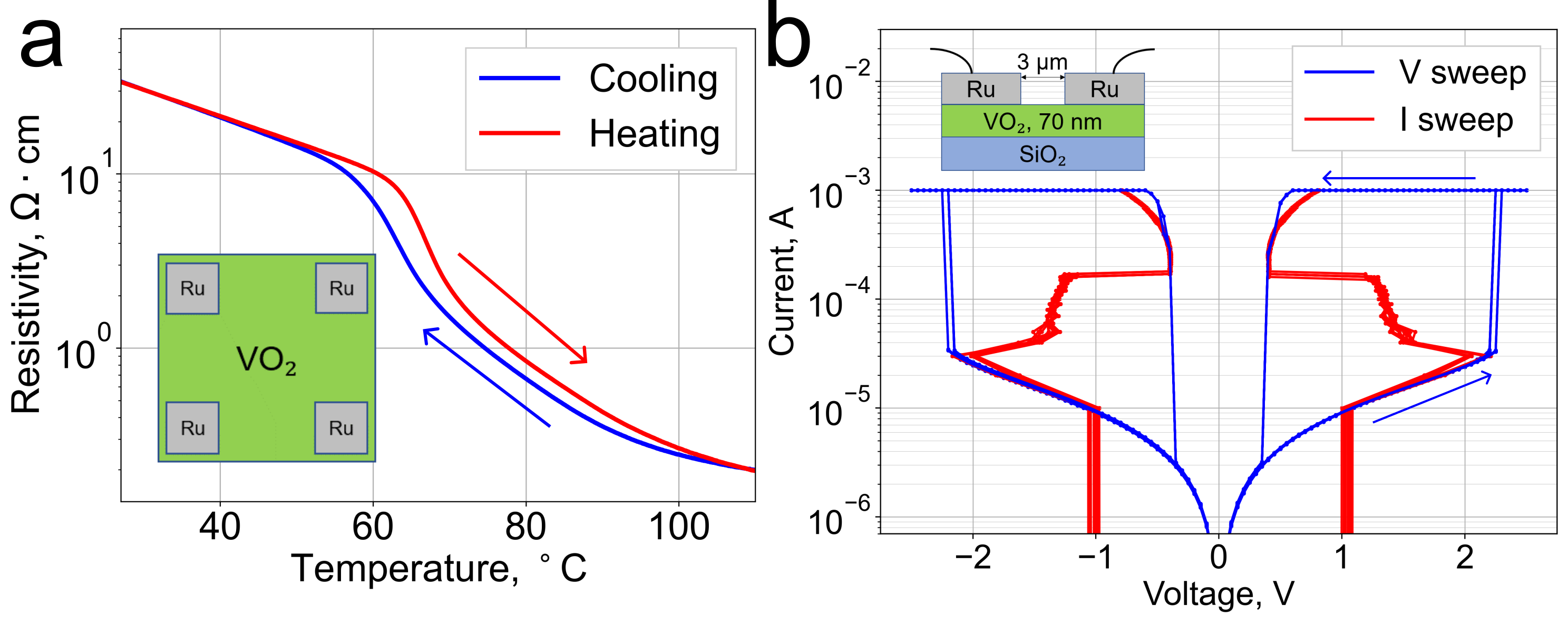}
\caption{a) Temperature dependence of resistivity of a VO$_2$ film measured in Van Der Pauw geometry. b) DC $I$$-$$V$ curves of Ru/VO$_2$/Ru lateral device: voltage sweep curves are drawn in blue (blue arrows indicate sweep direction), current sweep curves -- in red. Insets show schematics of the measuring methods and device structures.}
\label{trs_M6}
\end{center}
\end{figure}

\hspace{12pt} To unambiguously prove the presence of vanadium dioxide, suitable for neuromorphic applications, in our sample grown at optimized conditions, the presence of metal-insulator transition has to be established. To explore the electro-thermal properties of our VO$_2$ sample, temperature dependence of resistivity was measured (\textbf{fig.\ref{trs_M6}a}). One can observe the metal-insulator transition (MIT) at about $63~^\circ$C with a quite narrow ($\sim$$ 4~^\circ$C) hysteresis curve, elongated toward higher temperatures and retaining the residual hysteresis up to almost 100$~^{\circ}$C. This "blurred" $\rho(T)$ hysteresis curve can be explained by the distribution of transition temperatures and widths over the film volume \cite{hattori_investigation_2020, alzate_banguero_optical_2025}, which is plausible in the case of polycrystalline films. Also, a relatively high resistivity value of our films ($>300~\Omega~$cm at RT), in comparison to literature \cite{maher_highly_2024, varini_pulsed_2025}, appears to be profitable for neuromorphic applications, decreasing the energy consumption of artificial neurons.

To further test the eligibility of our VO$_2$ films for artificial neuron implementation, we measure current-voltage ($I$$-$$V$) curves of Ru/VO$_2$/Ru lateral resistive switches (fig.\ref{trs_M6}b). The devices demonstrate volatile resistive switching behavior: high-resistance state initially, abrupt jump into low-resistance state at a certain voltage value (the so called, "threshold" voltage), preservation of the low-resistance state down to a certain voltage of the same polarity (the so called, "hold" voltage), and automatic (with no need for manual reset) switching back to the high-resistance state. This phenomena occur as a result of Joule heating released due to the current flow, followed by local MIT in the VO$_2$ film \cite{woo_localized_2025}. Current compliance was set at 1$~$mA to prevent overheating and fast degradation of the device. Notably, the resistance ratio $R_{\text{HRS}}/R_{\text{LRS}}$ is larger than $10^2$ (taken at about 0.5$~$V), and the threshold voltage is quite low, yielding threshold electric field of less than $10^{-4}~$V/cm. Furthermore, by measuring $I$$-$$V$ curves in the current sweeping mode, we observe the presence of the negative differential resistance (NDR) region, which is crucial for the stabilization of self-oscillations and, consequently, realization of artificial neurons.

\section{Conclusion}\label{s:conclusion}

\hspace{12pt} To summarize, we have developed a method of targeted growth of vanadium dioxide thin films using reactive PLD from metallic V target at RT and subsequent annealing in vacuum. Formation of the necessary VO$_2$ phase occurs due to valence reduction and crystallization of the initially amorphous VO$_x$ thin film during vacuum annealing, which was directly confirmed by XPS measurements with \textit{in situ} annealing. By varying oxygen pressure during deposition, we have demonstrated the possibility to obtain different crystalline VO$_x$ phases and pure VO$_2$ in particular. Electrical measurements confirm the suitability of our VO$_2$ thin films for neuromorphic applications, with the properties of high initial resistance and low threshold voltage, which are promising for low power consumption of the devices. The proposed method has an additional advantage in the synthesis of vanadium dioxide without the need for substrate heating during growth, which simplifies the fabrication process and relaxes the equipment requirements.

\section*{Acknowledgments}

The work was performed using the equipment of the Center of Shared Research Facilities of Moscow Institute of Physics and Technology. This work was partially supported by the Ministry of Science and Higher Education of the Russian Federation (Agreement 075-03-2025-662 17.01.2025, project FSMG-2023-0006).

\section*{Conflict of Interest}

The authors declare no conflict of interest.

\section*{Data Availability Statement}

The data that support the findings of this study are available from the corresponding author upon reasonable request.

\bibliography{references}

\end{document}



\centering{\LARGE{Supporting Information}}
\begin{figure}[h]
\begin{center}
\includegraphics[width=\linewidth]{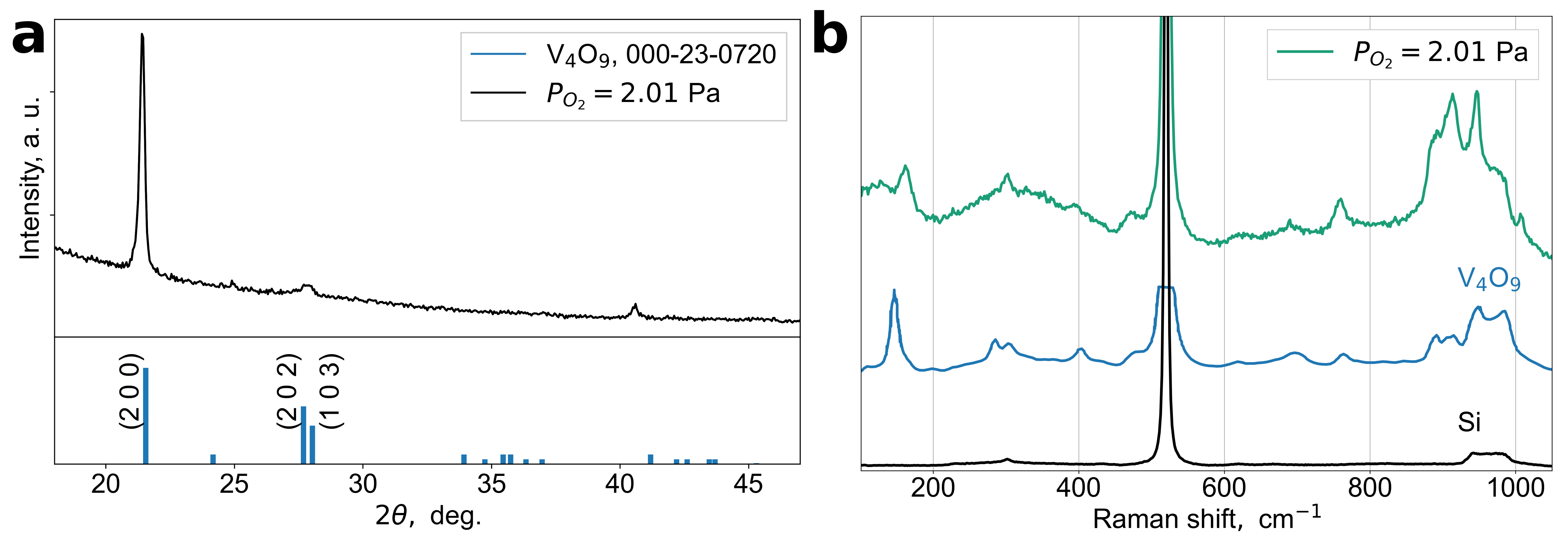}
\caption{XRD pattern (a) and Raman spectrum (b) of a sample grown with $2.01~$Pa oxygen pressure. The phase, matching the experimental measurements best, is V$_4$O$_9$.}
\label{v4o9}
\end{center}
\end{figure}

\begin{figure}[h]
\begin{center}
\includegraphics[width=\linewidth]{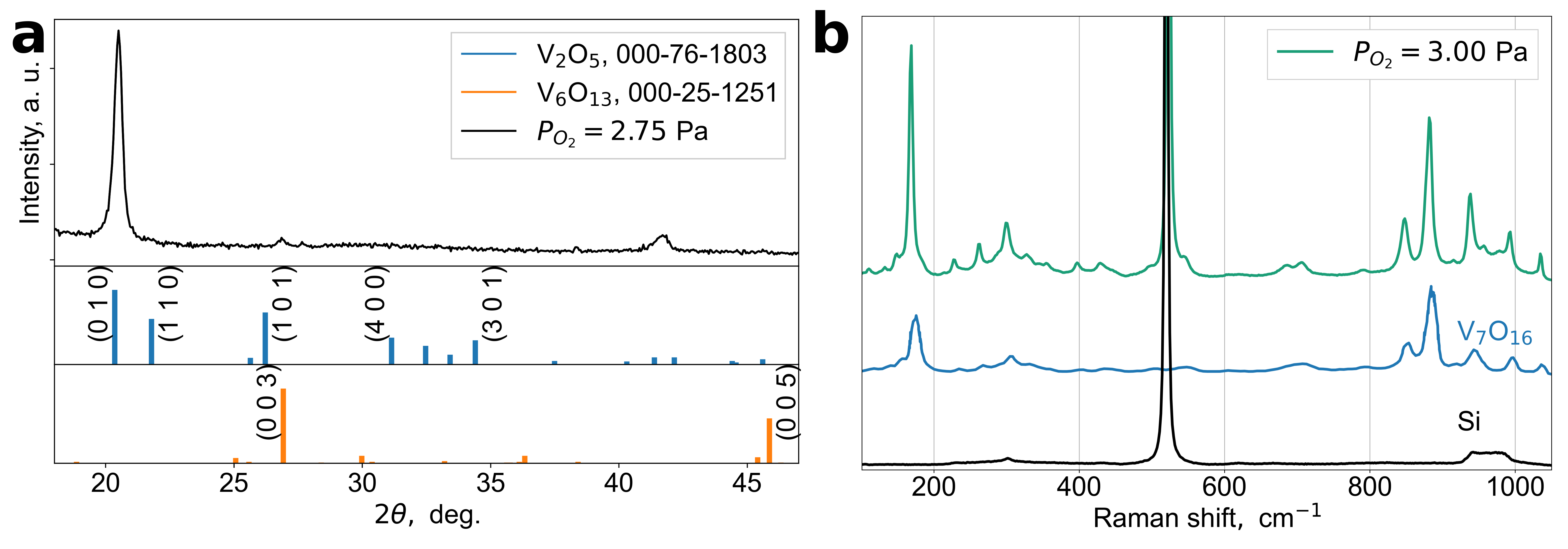}
\caption{a) XRD pattern of a sample grown with $2.75~$Pa oxygen pressure. Main peak correspond to a V$_2$O$_5$(010), also there is a small peak close to V$_6$O$_{13}$(003). b) Raman spectrum of a film grown with $3.00~$Pa oxygen pressure. One notes close similarity with a V$_7$O$_{16}$ reference spectrum.}
\label{other_phases}
\end{center}
\end{figure}